\documentclass[11pt,a4paper]{article}
\usepackage{jheppub}

\usepackage{graphicx}
\usepackage[utf8]{inputenc}
\usepackage{color}
\usepackage{bbm}
\usepackage{tensor}
\usepackage[all,cmtip]{xy}
\usepackage{dcolumn}
\usepackage{bm}

\makeatletter
\newcommand{\raisemath}[1]{\mathpalette{\raisem@th{#1}}}
\newcommand{\raisem@th}[3]{\raisebox{#1}{$#2#3$}}
\makeatother

\usepackage{tensor}
\usepackage{physics}
\usepackage{fontawesome}
\usepackage{slashed}
\usepackage{amsmath,amssymb,calc, amsthm,bbm, epsfig,psfrag, mathtools}
\usepackage[normalem]{ulem} 

\usepackage{latexsym,bm,amsfonts}
\usepackage{graphicx, enumerate}
 \usepackage[all,cmtip]{xy}
  \usepackage{float}
\allowdisplaybreaks

\newcommand{\Comment}[1]{{}}
\definecolor{darkblue}{rgb}{0.15,0.35,0.55}
\definecolor{reddish}{rgb}{0.65, 0.2, 0.2}

\newcommand{\be}{\begin{equation}}
\newcommand{\ee}{\end{equation}}
\newcommand{\bea}{\begin{equation}\begin{aligned}}
\newcommand{\eea}{\end{aligned}\end{equation}}

\newfont{\goth}{ygoth.tfm scaled 1200}                 

\def\1{{(1)}}
\def\2{{(2)}}
\def\3{{(3)}}

\newcommand {\cM}{{\cal M}}

\newcommand{\matter}{{\mathrm{M}}}

\newcommand{\field}{\bm{\varphi}}

\newcommand{\ba}{\begin{array}}
\newcommand{\ea}{\end{array}}

\def\double #1{#1{\hbox{\kern-2pt $#1$}}}

\newcommand{\bsubeq}{\begin{subequations}}
\newcommand{\esubeq}{\end{subequations}}

\def\tr{{\rm tr}}

\title{Solutions to the Ricci Flow via Einstein Field Equations}
\author[a]{Tommaso Morone,}
\author[a]{Roberto Tateo}

\affiliation[a]{Dipartimento di Fisica, Universit\`a di Torino and INFN, Sezione di Torino,\\
Via P. Giuria 1, 10125 Torino, Italy}

\emailAdd{tommaso.morone@unito.it}
\emailAdd{roberto.tateo@unito.it}
\abstract{
We investigate the relation between stress-tensor deformations of
matter theories coupled to gravity and the Ricci flow. We identify
the quadratic stress-tensor deformation whose associated metric
evolution, on solutions of the Einstein equations, coincides with the
Ricci-flow vector field. We show that this pointwise identification
does not in general extend to finite deformation parameter: the coupled
evolution of the metric and stress tensor must additionally preserve
the Einstein equations. We formulate this requirement as the invariance
of the Einstein constraint surface and derive the corresponding geometric
compatibility condition. We illustrate the resulting obstruction with
simple examples and exhibit the Bertotti--Robinson/Born--Infeld flow as
a nontrivial sector in which the correspondence is exact.
Finally, we introduce an Einstein-compatible completion of the
deformation which leaves its associated evolution of the metric
unchanged while modifying the stress-tensor evolution so that the
Einstein constraint surface is invariant by construction.
The completed evolution therefore generates genuine Ricci-flow
trajectories from arbitrary Einstein--matter initial data, whenever
the corresponding local evolution exists and is unique.
}

\begin{document}
\maketitle
\flushbottom
\section{Introduction}
The Ricci flow \cite{Ham:1982,chow2006ricci}, introduced by Hamilton in 1982, is a geometric evolution equation for Riemannian metrics with a central role in differential geometry and topology.
It evolves a metric according to its Ricci curvature and provides a natural mechanism for redistributing curvature across the manifold. Its
mathematical impact was highlighted by Perelman's proof of the
Poincar\'e conjecture
\cite{perelman2002entropy,perelman2003ricci}. 
Several generalisations have also been considered, including the Ricci--Bourguignon flow \cite{catino2017ricci}, in which the evolution equation contains an additional contribution proportional to the scalar curvature. Connections between the Ricci flow and physical systems, including black holes and nonlinear sigma models, have also been explored~\cite{Headrick:2006ti,Oliynyk:2005ak}.

In this paper we consider a one-parameter family of Lorentzian
manifolds $(\cM,g)_\tau$, with $\dim\cM=d\geq4$, and focus on the Ricci flow
\be\label{Ricflow}
    \dv{g_{ab}}{\tau}
    =
    -2R_{ab}\,,
    \qquad
    g_{ab}(\tau=0)=g_{0,ab}\,.
\ee
In Lorentzian signature we regard \eqref{Ricflow} as a geometric
evolution equation and do not assume the parabolic well-posedness
properties of the standard Riemannian Ricci flow.
Our aim is to investigate its relation with stress-tensor
deformations of matter theories coupled to gravity.

Stress-tensor deformations, and in particular $T\overline T$-like deformations, have been extensively studied in a variety of physical settings
\cite{Zamolodchikov:2004ce,Cavaglia:2016oda,Smirnov:2016lqw,
Jiang:2019epa,Taylor:2018xcy,Bonelli:2018kik,Conti:2018jho,
Ferko:2023sps,Ferko:2023wyi,Ebert:2024zwv,Ferko:2024zth}.
For a recent review and further references, see also
\cite{He:2025review}.
An important feature of these deformations is that they admit a characteristic description in terms of an auxiliary evolution in the space of metrics
\cite{Conti:2022egv,Morone:2024ffm,Hao:2024stt}. Recent developments have further emphasised metric-flow and gravitational formulations of stress-tensor deformations in arbitrary dimensions
\cite{Ran:2025holography,Li:2025emergent,Xie:2026gravitational}. This suggests a natural connection between stress-tensor deformations and geometric flows.

We consider a matter theory coupled to Einstein gravity and deform its matter action by the quadratic operator
\be\label{Tflow}
    \pdv{S_\tau^{\matter}}{\tau}
    =
    \frac12
    \int_\cM\mathrm{d}^dx\,\sqrt g\,
    \left[
        T_{ab}T^{ab}
        -
        \kappa T^2
    \right].
\ee
The corresponding auxiliary metric evolution is algebraic in the
stress-energy tensor. Using the Einstein equations, one finds that for
the distinguished value $\kappa=1/(d-2)$ this evolution coincides
pointwise with the Ricci-flow vector field. In this sense the Einstein equations provide a direct bridge between the stress-tensor deformation and the geometric evolution \eqref{Ricflow}. 
We will also show that, within the class of local algebraic
stress-tensor deformations considered here, the corresponding
quadratic operator is essentially singled out by this requirement.
There is, however, an important distinction between this instantaneous
identification and a genuine equivalence of finite flows. The Einstein
equations identify the auxiliary metric evolution with the Ricci-flow
vector field whenever they are satisfied, but the full coupled evolution need not preserve the Einstein equations. This is the central issue addressed in the present work. We regard the
Einstein equations as defining a constraint surface in the space of
metrics and stress-energy tensors and ask whether the characteristic
vector field preserves this surface. This separates pointwise matching
of the metric evolution from preservation of the gravitational constraint along the finite flow.

The latter requirement leads naturally to geometric compatibility
conditions. We derive the first obstruction and show that its vanishing expresses tangency of the characteristic vector field to the constraint
surface. This is only the first member of a hierarchy: tangency at the
initial point need not imply invariance for finite deformation parameter.

A small number of examples makes this distinction particularly
transparent. The Bertotti--Robinson solution provides a nontrivial case in which the correspondence is exact. In this background the stress-tensor deformation of Maxwell electrodynamics gives Born--Infeld electrodynamics, while the associated geometry follows the Ricci flow for finite deformation parameter. Related geometric
descriptions of nonlinear electrodynamics in terms of field-dependent metrics have also been developed in Ref.~\cite{Ferko:2026fielddependent}.
By contrast, Reissner--Nordstr\"om satisfies the scalar compatibility condition but fails the full tensorial test. A simple Friedmann--Lema\^{\i}tre--Robertson--Walker (FLRW) example goes one step further: it satisfies the complete first obstruction but fails at the
next order. 
Together, these examples show explicitly why pointwise matching, and
even first-order tangency, are not sufficient for a generic finite
correspondence.

This geometric formulation also suggests a simple way of restoring exact compatibility. For a characteristic generator
$\mathcal O[g,\mathbf T]$ we introduce the constraint-subtracted
combination
\be\label{OEintro}
    \mathcal O_{\rm E}[g,\mathbf T]
    =
    \mathcal O[g,\mathbf T]
    -
    \mathcal O[g,-\mathbf G]\,.
\ee
The subtraction leaves the associated evolution of the metric
unchanged, while modifying the stress-tensor evolution so that the
Einstein constraint surface is preserved.
Applied to the quadratic deformation associated with the Ricci
characteristic, this construction therefore generates genuine
Ricci-flow trajectories from arbitrary Einstein--matter initial data,
whenever the corresponding local evolution exists and is unique.

The geometric term appearing in the completion is directly related
to the first compatibility obstruction and admits a natural
perturbative interpretation as an infinitesimal metric displacement
proportional to the Ricci tensor.

The resulting picture is thus simple. The quadratic stress-tensor
deformation reproduces the Ricci-flow vector field on the Einstein
constraint surface, but its finite trajectories coincide with Ricci
flow only for compatible initial configurations. The completed
deformation instead builds preservation of the Einstein equations
into the characteristic evolution itself.

Although we concentrate here on the Ricci flow, the underlying idea
is considerably more general. Whenever a deformation of a matter
theory generates a characteristic vector field which agrees, on a
gravitational constraint surface, with a prescribed geometric flow,
one may separately study preservation of that constraint and construct
a corresponding constraint-subtracted completion. We expect this
viewpoint to be useful for other geometric flows and gravitational
systems as well.

The paper is organised as follows.
Section~\ref{charsec} reviews the characteristic
description of stress-tensor deformations and identifies the
quadratic operator associated with the Ricci characteristic.
Section~\ref{compatsec} formulates finite compatibility as
preservation of the Einstein constraint surface and derives the
corresponding geometric obstruction. 

Section~\ref{BRsec} discusses the exact
Bertotti--Robinson/Born--Infeld trajectory, while
Section~\ref{obssec} presents the Reissner--Nordstr\"om and
FLRW obstruction tests.
Section~\ref{OEsec} introduces the
Einstein-compatible completion and establishes preservation of the
Einstein equations under the completed characteristic flow. We
conclude in Section~\ref{conclusions} with a discussion of the
resulting picture and possible extensions.
Finally, Appendix~\ref{appgeom} collects the geometric compatibility
conditions and their higher-order extension, while
Appendix~\ref{apptests} contains the explicit obstruction tests for
the Reissner--Nordstr\"om and FLRW examples.
\section{Flows in the space of metrics}
\label{charsec}
In this section we summarise the characteristic description of
stress-tensor deformations and derive the coupled evolution equations
for the metric and stress-energy tensor that will be needed below.
Throughout, we use $g\equiv |\det g_{ab}|$. Consider a matter action
$S_\tau^{\matter}[g,\field]$ satisfying
\be\label{genflow}
    \pdv{S_\tau^{\matter}[g,\field]}{\tau}
    =
    \int_{\cM}\mathrm{d}^d x\,\sqrt{g}\,
    \mathcal O_\tau(\mathbf T)\,,
\ee
where $\mathcal O_\tau$ is a local algebraic functional of the
stress-energy tensor,
\be\label{Tdef}
    T_{\tau,ab}
    =
    \frac{2}{\sqrt{g}}\,
    \frac{\delta S_\tau^{\matter}}
         {\delta g^{ab}}\,.
\ee
With this sign convention, $T_{ab}$ is the negative of the
stress-energy tensor in the commonly used convention
$T^{\rm std}_{ab}=-(2/\sqrt{g})\,\delta S^{\matter}/\delta g^{ab}$.
We work in units in which $8\pi G_{\rm N}=1$. Accordingly, defining
the Einstein tensor by
\be
    G_{ab}\equiv R_{ab}-\frac12 Rg_{ab}\,,
\ee
the Einstein equations take the form
$\mathbf G=-\mathbf T$.
Unless otherwise stated, no derivatives of $T_{ab}$ are included in
$\mathcal O_\tau$.
Our curvature conventions are
\be
    R^a{}_{bcd}
    =
    \partial_c\Gamma^a_{db}
    -
    \partial_d\Gamma^a_{cb}
    +
    \Gamma^a_{ce}\Gamma^e_{db}
    -
    \Gamma^a_{de}\Gamma^e_{cb}\,,
    \qquad
    R_{bd}=R^a{}_{bad}\,,
\ee
and we define
\be
    \Delta
    \equiv
    g^{ab}\nabla_a\nabla_b\,, \qquad \nabla_a v^b = \partial_a v^b + \Gamma_{ac}^b v^c\,.
\ee
Thus, in Lorentzian signature, $\Delta$ denotes the covariant
d'Alembertian.

The deformation \eqref{genflow} admits a characteristic
description as an auxiliary evolution in the space of metrics
\cite{Conti:2022egv,Morone:2024ffm,Hao:2024stt},
\be\label{gflow}
    \dv{g_{ab}}{\tau}
    =
    2
    \frac{\partial\mathcal O_\tau}
         {\partial T_\tau^{ab}}\,,
    \qquad
    g_{ab}(0)=g_{0,ab}\,.
\ee
We refer to this auxiliary metric evolution as the metric
characteristic of the deformation.
It is convenient to introduce the mixed stress tensor
\be\label{mixedT}
    \mathbf T^a{}_b
    \equiv
    T^a{}_b
    =
    g^{ac}T_{cb}
\ee
and regard $\mathcal O_\tau$ as a class function of $\mathbf T$,
\be
    \mathcal O_\tau
    =
    \mathcal O_\tau
    \left(
        \tr\mathbf T,
        \tr\mathbf T^2,
        \ldots
    \right).
\ee
Defining
\be\label{Dmat}
    \mathbf D
    \equiv
    \frac{\partial\mathcal O_\tau}
         {\partial\mathbf T}\,,
\ee
one has $[\mathbf D,\mathbf T]=0$, and the metric characteristic becomes
\be\label{gmatrix}
    \mathbf g^{-1}\dot{\mathbf g}
    =
    2\mathbf D\,.
\ee
Here and below a dot denotes differentiation with respect to $\tau$.
\subsection{Evolution of the stress-energy tensor}
The metric characteristic alone is not sufficient to determine
whether a trajectory remains on shell. One must simultaneously follow
the stress-energy tensor. For a local algebraic deformation,
differentiating \eqref{Tdef} along the characteristic and using
\eqref{gflow} gives
\be\label{Tchar}
    \dot{\mathbf T}
    =
    -(\tr\mathbf D)\mathbf T
    +
    \left[
        \tr(\mathbf D\mathbf T)
        -
        \mathcal O_\tau
    \right]\mathbf 1
    \,.
\ee
Together, Eqs.~\eqref{gmatrix} and
\eqref{Tchar} constitute the coupled characteristic system for
the pair $(g,\mathbf T)$.

A useful consequence follows by taking the traceless part,
\be\label{Xdef}
    \mathbf X
    \equiv
    \mathbf T
    -
    \frac{1}{d}
    (\tr\mathbf T)\mathbf 1\,.
\ee
Equation~\eqref{Tchar} implies
\be\label{Xflow}
    \dot{\mathbf X}
    =
    -(\tr\mathbf D)\mathbf X\,,
\ee
and therefore
\be
    \mathbf X(\tau)
    =
    \alpha(\tau)\mathbf X(0)\,,
    \qquad
    \alpha(\tau)
    =
    \exp\left[
        -\int_0^\tau
        \mathrm{d}\tau'\,
        \tr\mathbf D(\tau')
    \right]\,.
\ee
Equivalently,
\be\label{Taffine}
    \mathbf T(\tau)
    =
    \alpha(\tau)\mathbf T_0
    +
    \beta(\tau)\mathbf 1
    \,.
\ee
Thus, along any regular characteristic trajectory, local algebraic
stress-tensor deformations preserve the eigenspaces and generalised
eigenspaces of the mixed stress tensor, together with their algebraic
and geometric multiplicities. In particular, if $\lambda_i(\tau)$
denote its eigenvalues,
\be
    \lambda_i(\tau)-\lambda_j(\tau)
    =
    \alpha(\tau)
    \big[\lambda_i(0)-\lambda_j(0)\big]\,.
\ee
This spectral rigidity will be useful below, but the main point for
our purposes is more basic: the characteristic evolution is a coupled
flow of both $g$ and $\mathbf T$, and compatibility with the Einstein
equations cannot be inferred from the metric equation alone.
\subsection{The Ricci-compatible quadratic operator}   
We now specialise to the quadratic deformation relevant for the
Ricci flow,
\be\label{Okappa}
    \mathcal O_\kappa
    =
    \frac12
    \left[
        \tr(\mathbf T^2)
        -
        \kappa(\tr\mathbf T)^2
    \right].
\ee
Its matrix derivative is
\be
    \mathbf D_\kappa
    =
    \mathbf T
    -
    \kappa T\,\mathbf 1\,,
    \qquad
    T\equiv\tr\mathbf T\,,
\ee
so that the metric characteristic reads
\be\label{gquad}
    \dot g_{ab}
    =
    2
    \left(
        T_{ab}
        -
        \kappa T g_{ab}
    \right).
\ee
Equation~\eqref{Tchar} similarly gives
\be\label{Tquad}
    \dot{\mathbf T}
    =
    -(1-d\kappa)T\,\mathbf T
    +
    \mathcal O_\kappa\,\mathbf 1\,.
\ee
For the distinguished value
\be\label{kappaR}
    \kappa_R
    =
    \frac{1}{d-2}\,,
\ee
we define
\be\label{OR}
    \mathcal O_R
    =
    \frac12
    \left[
        \tr(\mathbf T^2)
        -
        \frac{T^2}{d-2}
    \right].
\ee
The coupled characteristic equations become
\be\label{Rchars}
    \boxed{
    \begin{aligned}
        \mathbf g^{-1}\dot{\mathbf g}
        &=
        2
        \left(
            \mathbf T
            -
            \frac{T}{d-2}\mathbf 1
        \right),
        \\
        \dot{\mathbf T}
        &=
        \frac{2T}{d-2}\mathbf T
        +
        \mathcal O_R\,\mathbf 1 .
    \end{aligned}
    }
\ee
Suppose that at some value of the deformation parameter the pair
$(g,\mathbf T)$ satisfies the Einstein equations, 
\be\label{Einstein}
    \mathbf G=-\mathbf T\,.
\ee
Their trace gives
\be
    T
    =
    \frac{d-2}{2}R\,,
\ee
while trace reversal yields
\be\label{TRrev}
    \mathbf T
    -
    \frac{T}{d-2}\mathbf 1
    =
    -\mathbf R\,.
\ee
Consequently, at every point of the Einstein constraint surface,
the first equation in
\eqref{Rchars} becomes
\be\label{Ricinstant}
    \dot g_{ab}
    =
    -2R_{ab}
    \,.
\ee
The operator $\mathcal O_R$ is essentially singled out by this property.
Indeed, consider an arbitrary local algebraic class function
$\mathcal O(\mathbf T)$ and impose the Ricci-matching condition
universally in stress-tensor space, without restricting to a particular
matter sector or imposing additional algebraic relations on
$\mathbf T$. Using the Einstein trace-reversal relation, this requires
\be
    \frac{\partial\mathcal O}
         {\partial\mathbf T}
    =
    \mathbf T
    -
    \frac{\tr\mathbf T}{d-2}\mathbf 1\,.
\ee
Integration in stress-tensor space gives
\be\label{ORunique}
    \mathcal O(\mathbf T)
    =
    \frac12
    \left[
        \tr(\mathbf T^2)
        -
        \frac{(\tr\mathbf T)^2}{d-2}
    \right]
    +
    c
    =
    \mathcal O_R(\mathbf T)+c\,,
\ee
where $c$ is independent of $\mathbf T$. Thus, within the class of local algebraic stress-tensor functionals
considered here, and up to terms independent of $\mathbf T$ which do
not affect the metric characteristic, $\mathcal O_R$ is uniquely
singled out by the universal Ricci-matching condition in stress-tensor
space.

Equation~\eqref{Ricinstant} is the basic relation
underlying the correspondence considered in this paper. It is,
however, a pointwise statement: it applies whenever the Einstein
equations hold, but the coupled characteristic system
\eqref{Rchars} is defined independently of those
equations and need not preserve them.

The finite-flow problem is therefore distinct from the instantaneous
matching problem. A genuine Ricci-flow trajectory requires the
Einstein constraint surface to be invariant under the full coupled
characteristic evolution. We turn to this compatibility condition in
the next section.
\section{Einstein compatibility and geometric obstructions}
\label{compatsec}

We now address the main question underlying the relation between the
stress-tensor deformation and the Ricci flow. As discussed in the
previous Section, whenever the Einstein field equations are satisfied,
the metric characteristic generated by $\mathcal O_R$ obeys
\be
    \dv{g_{ab}}{\tau}=-2R_{ab}\,.
\ee
This pointwise identification of the metric characteristic with the
Ricci-flow vector field does not, however, guarantee that the
corresponding finite trajectories coincide. The coupled characteristic
evolution must also preserve the Einstein field equations.

To formulate this requirement, we introduce the mixed Einstein
constraint
\be\label{constraint}
    \mathbf C=\mathbf G+\mathbf T\,,
\ee
so that the Einstein equations define the constraint surface
$\mathbf C=0$ in the space of pairs $(g,\mathbf T)$. Let
$\mathcal V$ denote the characteristic vector field generated by the
coupled evolution of the metric and stress tensor. 

Exact compatibility requires the characteristic vector field to be tangent to the Einstein constraint surface at each of its points,
\be\label{tangent}
    \left.
    \mathcal L_{\mathcal V}\mathbf C
    \right|_{\mathbf C=0}
    =0\,.
\ee
This is an invariance condition on the full constraint surface, rather than merely a tangency condition at a particular initial configuration.

Here $\mathcal L_{\mathcal V}$ denotes differentiation along the
characteristic flow in field space, and should not be confused with a
space-time Lie derivative.

At a given Einstein–matter configuration, the corresponding tangency condition can be written entirely in geometric terms. On the Einstein constraint surface,
\be\label{TRrel}
    \mathbf T
    =
    -\mathbf R+\frac12R\,\mathbf 1\,,
    \qquad
    T=\frac{d-2}{2}R\,.
\ee
Comparing the characteristic evolution of the stress tensor with the
Ricci-flow evolution of the curvature gives the tensorial condition
\bea\label{firstobs}
    \mathfrak K^a{}_b
    \equiv{}&
    \Delta R^a{}_b
    +
    2R^a{}_{cbd}R^{cd}
    -
    RR^a{}_b
    \\
    &-
    \left(
        \frac12\Delta R
        +
        \frac12R_{cd}R^{cd}
        -
        \frac14R^2
    \right)\delta^a_b
    =
    0\,.
\eea
We refer to $\mathfrak K^a{}_b$ as the first geometric obstruction.
Its detailed derivation, including the separate trace and traceless
compatibility conditions, is given in
Appendix~\ref{appfirst}.

Taking the trace of \eqref{firstobs} gives
\be\label{scalarobs}
    (d-2)\Delta R
    +
    (d-4)
    \left(
        R_{ab}R^{ab}
        -
        \frac12R^2
    \right)
    =
    0\,.
\ee
In four dimensions this reduces to the particularly simple condition
\be\label{harmonicR}
    \Delta R=0\,.
\ee
Thus the first compatibility test already imposes a differential
constraint on the curvature of the initial Einstein--matter
configuration.
The condition
\be\label{firsttest}
    \mathfrak K^a{}_b=0
\ee
has a precise geometric meaning: it states that the characteristic
vector field is tangent to the Einstein constraint surface at the
given configuration. It is therefore necessary for finite
Ricci-flow compatibility, but is not sufficient in general. A
trajectory may satisfy \eqref{firsttest} initially and leave
the constraint surface at higher orders in the deformation parameter.

Successive compatibility conditions are obtained by differentiating
the Einstein constraint repeatedly along the characteristic flow. If
\be
    \mathbf C_0\equiv\mathbf C\,,
\ee
we define
\be\label{hierarchy}
    \mathbf C_{n+1}
    =
    \mathcal L_{\mathcal V}\mathbf C_n\,,
    \qquad
    n=0,1,2,\ldots\,.
\ee
For an analytic characteristic trajectory starting on the Einstein
constraint surface, exact compatibility requires
\be\label{allorders}
    \left.\mathbf C_n\right|_{\tau=0}=0\,,
    \qquad
    n=0,1,2,\ldots\,.
\ee
Conversely, within the analytic class, the vanishing of the full
hierarchy at the initial point implies local preservation of the
constraint through the corresponding Taylor expansion. The construction of this hierarchy is discussed in
Appendix~\ref{highercompat}.

The trajectory-independent counterpart of this hierarchy is the
intrinsic invariance condition \eqref{tangent} on the full Einstein
constraint surface. This condition implies the hierarchy above for
every Einstein--matter initial configuration.
A useful sufficient realisation is a closure relation of the form
\be\label{constraintclosure}
    \mathcal L_{\mathcal V}\mathbf C
    =
    \mathcal A[\mathbf C]\,,
\ee
where $\mathcal A$ is a local linear operator. Such a closure relation
makes the Einstein constraint surface invariant and automatically
implies the higher compatibility conditions.

The following Sections illustrate these possibilities through a small
number of representative examples. We first discuss a nontrivial
sector in which the correspondence is exact, and then turn to
configurations for which compatibility fails either at the first
obstruction or at higher order.
\section{An exact example: Bertotti--Robinson and Born--Infeld}
\label{BRsec}
We first illustrate the compatibility conditions with a nontrivial
sector in which the correspondence is exact. Consider the
four-dimensional Bertotti--Robinson geometry
\be\label{BRmetric}
    \mathrm d s_0^2
    =
    Q^2\,\mathrm d s^2_{\mathrm{AdS}_2}
    +
    Q^2\,\mathrm d\Omega_2^2\,,
\ee
where the two factors have equal curvature radii. The mixed Ricci
endomorphism has the spectrum
\be\label{BRspec}
    \operatorname{spec}(\mathbf R_0)
    =
    \left\{
        -\frac1{Q^2},-\frac1{Q^2},
        \frac1{Q^2},\frac1{Q^2}
    \right\},
\ee
and therefore the Ricci scalar vanishes, $R_0=0$.
The geometry is supported by Maxwell electrodynamics,
\be\label{Maxwell}
    S_0^{\matter}[g_0,A]
    =
    -\frac14
    \int_{\cM}\mathrm d^4x\,\sqrt{g_0}\,
    F_{ab}F^{ab}\,,
    \qquad F=\mathrm dA\,.
\ee
The corresponding stress-energy tensor is traceless and has two
doubly-degenerate eigenvalues of opposite sign, as required by the
Einstein equations.
For a direct product of two two-dimensional Einstein spaces, a space-time independent rescaling of each factor leaves the corresponding covariant Ricci tensor unchanged. Indeed, a constant rescaling leaves the Levi-Civita connection, and hence the covariant Ricci tensor, unchanged.

The Ricci-flow equation can therefore be
integrated directly. Starting from \eqref{BRmetric}, one obtains
\be\label{BRflow}
    g(\tau)
    =
    \left(1+\frac{2\tau}{Q^2}\right)
    Q^2\,\mathrm d s^2_{\mathrm{AdS}_2}
    +
    \left(1-\frac{2\tau}{Q^2}\right)
    Q^2\,\mathrm d\Omega_2^2\,.
\ee
Equivalently, the running squared radii are
\be\label{BRradii}
    L_{\mathrm{AdS}}^2(\tau)=Q^2+2\tau\,,
    \qquad
    L_\mathrm{S}^2(\tau)=Q^2-2\tau\,.
\ee
For positive deformation parameter, the solution remains regular
within this product ansatz for
\be
    0\leq \tau < \frac{Q^2}{2}\,,
\ee
where the upper endpoint corresponds to the collapse of the
$\mathrm{S}^2$ factor.
The mixed Ricci eigenvalues consequently evolve as
\be\label{rsevol}
    r(\tau)=-\frac{1}{Q^2+2\tau}\,,
    \qquad
    s(\tau)=\frac{1}{Q^2-2\tau}\,.
\ee
The same trajectory is generated by the stress-tensor deformation.
In four dimensions,
\be
    \mathcal O_R
    =
    \frac12
    \left[
        \tr(\mathbf T^2)-\frac12T^2
    \right].
\ee
For a stress tensor with spectrum $\{u,u,v,v\}$, one has
\be
    \mathcal O_R=-2uv\,,
\ee
and the characteristic equation reduces to
\be\label{BRstress}
    \dot u=2u^2\,,
    \qquad
    \dot v=2v^2\,.
\ee
With the Bertotti--Robinson initial values
\[
    u(0)=\frac{1}{Q^2}
    \quad (\mathrm{AdS}_2\text{ block}),
    \qquad
    v(0)=-\frac{1}{Q^2}
    \quad (\mathrm{S}^2\text{ block}),
\]
these equations give
\be
    u(\tau)=\frac{1}{Q^2-2\tau}\,,
    \qquad
    v(\tau)=-\frac{1}{Q^2+2\tau}\,.
\ee
These are precisely the stress-tensor eigenvalues required by the
Einstein equations for the Ricci eigenvalues in \eqref{rsevol}.
Indeed, since
\be
    R=2(r+s)\,,
\ee
trace reversal gives
\be
    T_{\mathrm{AdS}_2}
    =
    -r+\frac12R=s\,,
    \qquad
    T_{\mathrm{S}^2}
    =
    -s+\frac12R=r\,.
\ee
Since the two eigenspaces are preserved by the characteristic evolution
and the corresponding eigenvalues coincide with those required by the
Einstein equations, the Einstein constraint is preserved for finite
$\tau$.

The matter flow has a familiar interpretation. With the normalisation
of the deformation parameter adopted here, the quadratic stress-tensor
deformation maps Maxwell theory to Born--Infeld electrodynamics
\cite{Conti:2018jho}.
The deformed action may then be written as
\be\label{BIaction}
    S_\tau^{\matter}[g,A]
    =
    \frac{1}{4\tau}
    \int_{\cM}\mathrm d^4x
    \left[
        \sqrt g
        -
        \sqrt{
        \left|
        \det\left(
            g_{ab}+\sqrt{4\tau}\,F_{ab}
        \right)
        \right|}
    \right].
\ee
Introducing
\be
    X=\frac14F_{ab}F^{ab}\,,
    \qquad
    Y=\frac14F_{ab}\widetilde F^{ab}\,,
\ee
one obtains
\be\label{BILag}
    \mathcal L_{\rm BI}(X,Y;\tau)
    =
    \frac{1-
        \sqrt{1+8\tau X-16\tau^2Y^2}}
         {4\tau}\,,
\ee
which reduces to Maxwell theory as $\tau\to0$.

Thus the Bertotti--Robinson solution provides a nontrivial finite
realisation of the correspondence: the matter theory evolves from
Maxwell to Born--Infeld electrodynamics, while the geometry follows
the exact Ricci-flow trajectory \eqref{BRradii}. The example
also makes clear that exact compatibility is a property of the
combined Einstein--matter evolution, rather than of the matter
deformation alone.
\section{Examples and obstructions}
\label{obssec}
The Bertotti--Robinson example shows that the characteristic and Ricci
flows can remain compatible for finite values of the deformation parameter. The correspondence is not generic, however. We now give two complementary examples: Reissner--Nordstr\"om fails already at the first tensorial
obstruction, whereas a radiation-like FLRW configuration passes the
complete first test but fails at the next order.

\subsection{Reissner--Nordstr\"om geometry}
\label{RNsec}

A particularly instructive example is provided by the
Reissner--Nordstr\"om solution of four-dimensional Einstein--Maxwell
theory. This case is especially relevant because Maxwell
electrodynamics is mapped into Born--Infeld electrodynamics by the
same stress-tensor deformation discussed in
Section~\ref{BRsec}. Nevertheless, the corresponding deformation of
a Reissner--Nordstr\"om solution does not define a Ricci-flow
trajectory.

For this geometry,
\be
    R=0\,,
\ee
so the scalar compatibility condition \eqref{harmonicR} is
automatically satisfied. The distinction appears at the tensorial
level. A direct evaluation of the first geometric obstruction gives
\be\label{RNobs}
    \mathfrak K^a{}_b
    =
    \frac{4Q^2 f(r)}{r^6}
    \operatorname{diag}(-3,-1,2,2)
    \,,
\ee
where
\be
    f(r)=1-\frac{2M}{r}+\frac{Q^2}{r^2}\,.
\ee
The calculation is given in
Appendix~\ref{appRN}. Although \eqref{RNobs} is traceless, as required by
$R=0$, it does not vanish identically for $Q\neq0$.

Thus, although the metric characteristic agrees with the Ricci-flow
vector field at the initial Einstein--Maxwell configuration, the full
characteristic vector field is not tangent to the Einstein constraint
surface. In particular, the
Maxwell-to-Born--Infeld deformation does not by itself turn the
Reissner--Nordstr\"om family into a Ricci-flow trajectory.

\subsection{A cosmological higher-order obstruction}
\label{FLRWsec}

The first geometric obstruction is necessary but not sufficient.
This can be seen explicitly in a spatially flat four-dimensional
FLRW geometry with
\be\label{FLRWseed}
    a(t)=a_0t^{1/2}\,.
\ee
For this configuration,
\be
    R=0\,,
    \qquad
    R^a{}_b
    =
    \frac{1}{4t^2}
    \operatorname{diag}(-3,1,1,1)\,,
\ee
and direct substitution into \eqref{firstobs} gives
\be\label{FLRWfirst}
    \mathfrak K^a{}_b=0
    \,.
\ee
Thus the characteristic vector field is tangent to the Einstein
constraint surface at the initial configuration. The tangency does not persist. Expanding the coupled metric and
stress-tensor characteristics in $\tau$ about the initial
configuration, one finds that the Einstein equations are satisfied
through first order but fail at second order:
\be\label{FLRWfail}
    \mathbf C_0=0\,,
    \qquad
    \mathbf C_1=0\,,
    \qquad
    \mathbf C_2\neq0
    \,.
\ee
The explicit expansion is given in
Appendix~\ref{appFLRW}.

The Reissner--Nordstr\"om and FLRW examples therefore probe two
different failures of finite compatibility. The former violates the
first tensorial condition, while the latter satisfies it but fails at
higher order. Together they show why the pointwise relation
$\dot g_{ab}=-2R_{ab}$ cannot by itself be promoted to an equality of
finite trajectories.
\section{An Einstein-compatible completion}
\label{OEsec}
The previous analysis shows that the metric evolution itself is not
the source of the obstruction. Whenever the Einstein equations are
satisfied, $\mathcal O_R$ generates
\be
    \dot g_{ab}=-2R_{ab}\,.
\ee
The problem is instead that the accompanying evolution of the
stress-energy tensor does not generically preserve the Einstein
constraint surface. This suggests modifying the deformation without
changing the associated metric evolution.

Let $\mathcal O[g,\mathbf T]$ be a local algebraic characteristic
generator of the type considered above. We define its
Einstein-compatible completion by

\be\label{OEgen}
    \boxed{
    \mathcal O_{\rm E}[g,\mathbf T]
    =
    \mathcal O[g,\mathbf T]
    -
    \mathcal O[g,-\mathbf G]
    }\,.
\ee
The subtraction term depends on the metric but not on the independent
stress-energy tensor. It therefore leaves the associated metric
evolution unchanged. Moreover,
\be
    \left.
    \mathcal O_{\rm E}
    \right|_{\mathbf T=-\mathbf G}
    =
    0\,.
\ee
The vanishing of $\mathcal O_{\rm E}$ on the constraint surface is not
by itself sufficient to guarantee that the surface is preserved. To
establish this stronger property, we now turn to the characteristic
evolution at the canonical level. Let $q^I$ and $p_I$ denote
configuration and conjugate variables. We consider a constraint surface
of gradient type,
\be
    p_I=p_{E,I}(q)\,,
    \qquad
    p_{E,I}(q)=\frac{\partial W_E(q)}{\partial q^I}\,,
\ee
as appropriate to the variational constraint relevant below. In
particular,
\be
\label{pp}
    \frac{\partial p_{E,J}}{\partial q^I}
    =
    \frac{\partial p_{E,I}}{\partial q^J}\,.
\ee
This condition holds automatically when the constraint surface is defined by a variational principle, since $p_{E,I}$ is then the gradient of a well-defined functional $W_E$, and mixed partial derivatives commute. The analogue of \eqref{OEgen} is
\be\label{HE}
    H_E(q,p)
    =
    H(q,p)-H\bigl(q,p_E(q)\bigr)\,.
\ee
Since the second term is independent of the independent momenta,
the configuration characteristic is unchanged,
\be
    \dot q^I
    =
    \frac{\partial H}{\partial p_I}\,.
\ee
On the constraint surface, the conjugate equation gives
\be
    \left.
    \dot p_I
    \right|_{p=p_E}
    =
    \left.
    \frac{\partial H}{\partial p_J}
    \right|_{p=p_E}
    \frac{\partial p_{E,J}}{\partial q^I}
    =
    \dot q^J
    \frac{\partial p_{E,I}}{\partial q^J}
    =
    \frac{\mathrm d}{\mathrm d\tau}
    p_{E,I}(q(\tau))\,,
\ee
and hence
\be\label{Hconstraint}
    \left.
    \frac{\mathrm d}{\mathrm d\tau}
    \left(
        p_I-p_{E,I}(q)
    \right)
    \right|_{p=p_E}
    =
    0\,.
\ee
Thus the completed characteristic vector field is tangent to the
constraint surface at every point of that surface.

For the gravitational problem, the gradient condition used above
follows directly from the variational formulation of the Einstein
equations. In the units $8\pi G_{\rm N}=1$ adopted above, the
Einstein--Hilbert action reads
\be
    S_{\rm EH}[g]
    =
    \frac12
    \int_{\cM}\mathrm{d}^d x\,\sqrt{g}\,R\,.
\ee
The canonical momentum is the variational momentum density associated
with the matter functional, while the Einstein equations identify its
constraint value with minus the corresponding variational derivative
of the Einstein--Hilbert functional.

The role of $W_E$ is therefore played by $-S_{\rm EH}$, up to
boundary terms. Choosing the inverse metric as configuration variable,
the corresponding variational momentum densities are
\be
    p_{ab}
    \equiv
    \frac{\delta S^{\matter}}{\delta g^{ab}}
    =
    \frac{\sqrt{g}}{2}\,T_{ab}\,,
    \qquad
    p_{E,ab}
    \equiv
    -\frac{\delta S_{\rm EH}}{\delta g^{ab}}
    =
    -\frac{\sqrt{g}}{2}\,G_{ab}\,,
\ee
with the normalisation adopted here. Hence $p_{ab}=p_{E,ab}$ is
equivalent to $\mathbf T=-\mathbf G$.

In the field-theory setting, the symmetry condition \eqref{pp}
corresponds to the symmetry of the second functional derivative,
assuming boundary conditions such that the relevant boundary terms
vanish.
Notice that $\mathbf T$ is not itself the canonical momentum:
the relation between the two involves the metric and the volume
density, as displayed above. The constraint-subtraction argument is
therefore formulated at the canonical level and only subsequently
translated into stress-tensor variables.

Applying the preceding argument at the canonical level and then
translating back to stress-tensor variables gives
\be\label{OEtangent}
    \left.
    \frac{\mathrm d}{\mathrm d\tau}
    \left(
        \mathbf G+\mathbf T
    \right)
    \right|_{\mathbf G+\mathbf T=0}
    =
    0\,.
\ee
Consequently, if
\be
    \mathbf G[g_0]+\mathbf T_0=0\,,
\ee
then
\be\label{Einpres}
    \mathbf G[g(\tau)]+\mathbf T(\tau)=0
\ee
throughout the interval in which the local characteristic evolution
is well-posed and unique.
The preservation of the Einstein constraint also guarantees the
covariant conservation of the stress-energy tensor along the
completed trajectory. Indeed, the contracted Bianchi identity implies
\be\label{Tcons}
    \nabla^a T_{ab}(\tau)=0\,.
\ee
Thus the completed characteristic evolution preserves both the
Einstein equations and the associated stress-energy conservation law.

For the quadratic Ricci-compatible operator we have
\be
    \mathcal O_R(\mathbf T)
    =
    \frac12
    \left[
        \tr(\mathbf T^2)
        -
        \frac{(\tr\mathbf T)^2}{d-2}
    \right]\,.
\ee
Since $\mathcal O_R$ is quadratic,
$\mathcal O_R(-\mathbf G)=\mathcal O_R(\mathbf G)$, and the
geometric subtraction is
\be
    \mathcal O_R(\mathbf G)
    =
    \frac12
    \left(
        R_{ab}R^{ab}
        -
        \frac12 R^2
    \right)\,.
\ee
The completed operator therefore takes the explicit form
\be\label{OEform}
    \mathcal O_{\rm E}
    =
    \frac12
    \left[
        T_{ab}T^{ab}
        -
        \frac{T^2}{d-2}
    \right]
    -
    \frac12
    \left(
        R_{ab}R^{ab}
        -
        \frac12R^2
    \right)
    \,.
\ee
Since the geometric term does not depend on the stress-energy tensor treated as an independent variable, the metric characteristic remains
\be
    \dot g_{ab}
    =
    2
    \left(
        T_{ab}
        -
        \frac{T}{d-2}g_{ab}
    \right).
\ee
The corresponding off-shell completed stress-tensor characteristic is
\be
    \dot{\mathbf T}
    =
    \frac{2T}{d-2}\,\mathbf T
    +
    \mathcal O_R(\mathbf T)\,\mathbf 1
    -
    \mathfrak K[g]\,.
\ee
Here $\mathfrak K[g]$ denotes the geometric tensor defined in
\eqref{firstobs}, evaluated for a generic metric; its origin from the
geometric subtraction is made explicit below using \eqref{I2var}.
Unlike the original algebraic characteristic system, the completed
evolution is no longer algebraic in the metric, since $\mathfrak K[g]$
contains derivatives of the curvature.

Equation~\eqref{Einpres} allows the Einstein
equations to be imposed at every value of $\tau$, and hence
\be\label{Ricexact}
    \boxed{
    \dot g_{ab}
    =
    -2R_{ab}[g(\tau)]
    }\,.
\ee
Thus, for arbitrary Einstein--matter initial data, the completed
characteristic generates a genuine Ricci flow throughout the interval
in which the corresponding local characteristic evolution exists and
is unique, rather than only for the special compatible sectors
selected by the original deformation.

For sectors in which the original deformation is already exactly
compatible, such as Bertotti--Robinson, the completion is not needed
to enforce preservation of the corresponding trajectory. Along such
a trajectory the first obstruction vanishes identically, so that the
original and completed characteristic vector fields coincide. For
generic Einstein--matter initial data, instead, the completion modifies
the stress-tensor evolution so as to preserve the Einstein constraint.

The structure of the completion becomes particularly transparent in
terms of the Einstein constraint
\be
    \mathbf C=\mathbf G+\mathbf T\,.
\ee
Writing
\be
    \mathbf T=-\mathbf G+\mathbf C
\ee
in \eqref{OEform}, one finds
\be\label{OEconstraint}
    \mathcal O_{\rm E}
    =
    -R^a{}_b C^b{}_a
    +
    \frac12
    \left[
        \tr(\mathbf C^2)
        -
        \frac{C^2}{d-2}
    \right]
    \,,
\ee
where $C=\tr\mathbf C$. Hence
\be
    \left.
    \mathcal O_{\rm E}
    \right|_{\mathbf C=0}
    =
    0\,,
\ee
while
\be
    \left.
    \frac{\partial\mathcal O_{\rm E}}
         {\partial T^a{}_b}
    \right|_{\mathbf C=0}
    =
    -R^b{}_a\,.
\ee
The completed deformation therefore vanishes on the Einstein
constraint surface while its metric characteristic remains
nontrivial:
\be\label{OEonshell}
    \left.
    \mathcal O_{\rm E}
    \right|_{\mathbf C=0}
    =
    0\,,
    \qquad
    \left.
    \dot g_{ab}
    \right|_{\mathbf C=0}
    =
    -2R_{ab}\,.
\ee
In this sense, $\mathcal O_{\rm E}$ has a constraint-like structure:
it vanishes on the Einstein constraint surface, while its derivatives
generate a nontrivial characteristic vector field that is tangent to
that surface.
There is also a direct relation between the geometric subtraction and
the first compatibility obstruction. Consider
\be\label{I2}
    I_2[g]
    =
    \int_{\cM}\mathrm d^dx\,\sqrt{g}\,
    \left(
        R_{ab}R^{ab}
        -
        \frac12R^2
    \right).
\ee
Up to boundary terms,
\be\label{I2var}
    \frac{1}{\sqrt g}
    \frac{\delta I_2}{\delta g^{ab}}
    =
    \mathfrak K_{ab}\,,
\ee
where $\mathfrak K_{ab}$ is precisely the first geometric obstruction
defined in Eq.~\eqref{firstobs}, with the curvature conventions
adopted throughout this paper.
Since the geometric term
subtracted in \eqref{OEform} is one half of the
integrand in \eqref{I2}, its
contribution to the conjugate characteristic equation is
\be
    -2\,
    \frac{1}{\sqrt g}
    \frac{\delta}{\delta g^{ab}}
    \int_{\cM}\mathrm d^dx\,\sqrt g\,
    \mathcal O_R(\mathbf G)
    =
    -\mathfrak K_{ab}\,.
\ee
At the infinitesimal level, the completion therefore cancels the
inhomogeneous geometric source responsible for the first departure of
the original characteristic from the Einstein constraint surface. The first obstruction is the
infinitesimal manifestation of the same constraint-subtraction
mechanism.

Finally, the curvature-squared subtraction has a simple perturbative
interpretation. Under the infinitesimal inverse-metric redefinition
\be
    g^{ab}
    \longrightarrow
    g^{ab}-\epsilon R^{ab}\,,
\ee
the Einstein--Hilbert action changes, up to boundary terms, by
\be
    \delta S_{\rm EH}
    =
    -\frac{\epsilon}{2}
    \int_{\cM}\mathrm d^dx\,\sqrt g\,
    \left(
        R_{ab}R^{ab}
        -
        \frac12R^2
    \right).
\ee
On the Einstein constraint surface, the corresponding variation of
the matter action contains
\be
    -\frac12T_{ab}R^{ab}
    =
    \frac12
    \left(
        T_{ab}T^{ab}
        -
        \frac{T^2}{d-2}
    \right)
    =
    \mathcal O_R(\mathbf T)\,.
\ee
The two terms entering $\mathcal O_{\rm E}$ may therefore be viewed
as the gravitational and matter responses to an infinitesimal metric
displacement proportional to the Ricci tensor.
Indeed, on the Einstein constraint surface their sum gives
\be
    \delta\!\left(S_{\rm EH}+S^{\matter}\right)
    =
    \epsilon
    \int_{\cM}\mathrm{d}^d x\,\sqrt{g}\,
    \mathcal O_{\rm E}
    =
    0\,.
\ee
This observation does not imply that the completed deformation should
be interpreted as a finite field redefinition, nor that
\eqref{OEform} must be regarded as a fundamental
higher-derivative gravitational action. Here its role is instead that
of a constraint-preserving characteristic completion, with the
field-redefinition argument providing a natural perturbative
interpretation of the curvature-squared subtraction.
\section{Discussion and conclusions}
\label{conclusions}
In this work we have explored the relation between stress-tensor
deformations of matter theories coupled to gravity and the Ricci flow. The basic correspondence is remarkably direct: for the
distinguished quadratic stress-tensor deformation, the metric
characteristic coincides with the Ricci-flow vector field whenever
the Einstein equations are satisfied. The key point is that this identification is intrinsically pointwise. A genuine
finite correspondence additionally requires the characteristic
evolution to preserve the Einstein equations.

This observation leads to a geometric formulation of the problem in
terms of the Einstein constraint surface. Its preservation gives rise
to compatibility conditions, the first of which is encoded in the
tensor $\mathfrak K_{ab}$ derived in
Eq.~\eqref{firstobs}. The examples considered in this paper
illustrate the different possibilities in a simple way.
Bertotti--Robinson provides a nontrivial exact trajectory, whereas
Reissner--Nordstr\"om fails already at the first obstruction. The
FLRW example shows that even the vanishing of the complete first
obstruction does not in general guarantee finite compatibility.

We have also shown that exact compatibility can be enforced by a simple
constraint-subtraction prescription. The Einstein-compatible completion
leaves the associated metric evolution unchanged while modifying the
stress-tensor evolution so that the Einstein constraint surface is
invariant.

The same construction also clarifies the geometric origin of the first obstruction: the curvature term entering the completion has a metric variation which
precisely cancels the source responsible for driving the original
characteristic away from the Einstein surface.

The resulting distinction is conceptually useful. For the original
stress-tensor deformation, exact Ricci-flow evolution occurs only for
those Einstein--matter configurations whose coupled evolution preserves
the Einstein constraint. For the completed deformation, this
preservation is instead built into the construction. These provide two
complementary ways of relating matter deformations to geometric flows.

We expect that this viewpoint extends considerably beyond the
specific Ricci-flow example studied here.

The essential ingredients
are a characteristic evolution, a gravitational constraint surface,
and a geometric flow obtained by restricting the characteristic
vector field to that surface.
Separating the matching of the vector fields from the preservation of the constraint provides a natural strategy for studying analogous correspondences in other settings. Constraint subtraction then offers a systematic way of enforcing compatibility when it is not automatic.

It would be interesting to explore this framework for other geometric
flows of relevance in physics and mathematics, as well as in
different dimensions, signatures and gravitational theories. A more
systematic investigation of the special geometries for which the
uncompleted deformation is already exact is another natural
direction. We leave these questions for future work.
\acknowledgments

We thank Paolo Aschieri, Hossain Babaei-Aghbolagh, Marco Billò,
Nicolò Brizio, Christian Ferko, Song He, Jue Hou, Hao Ouyang,
Nicolò Petri, Gabriele Tartaglino-Mazzucchelli and Evangelos
Tsolakidis for helpful discussions. We are particularly grateful to Edward Witten for insightful comments that helped us clarify the distinction between the pointwise identification of the flows and their finite compatibility. We are also very grateful to Marco Meineri for valuable feedback and encouragement throughout this research.

The authors received partial support from the INFN project
``Statistical Field Theory (SFT)'', and the PRIN (Progetti di
Rilevante Interesse Nazionale) Project No.~2022ABPBEY,
``Understanding quantum field theory through its deformations'',
funded by the Italian Ministry of University and Research.

\appendix
\section{Geometric compatibility conditions}
\label{appgeom}
In this appendix we collect the main technical steps entering the
compatibility analysis of Section~\ref{compatsec}.

\subsection{First geometric obstruction}
\label{appfirst}

On the Einstein constraint surface,
\be
    \mathbf T
    =
    -\mathbf R+\frac12R\,\mathbf 1\,,
    \qquad
    T=\frac{d-2}{2}R\,.
\ee
For the Ricci-compatible quadratic deformation, the characteristic
equation for the mixed stress-energy tensor is
\be
    \dot{\mathbf T}
    =
    \frac{2T}{d-2}\mathbf T
    +
    \mathcal O_R\,\mathbf 1\,,
\ee
with
\be
    \mathcal O_R
    =
    \frac12
    \left[
        \tr(\mathbf T^2)
        -
        \frac{T^2}{d-2}
    \right].
\ee
On the Einstein constraint surface one finds
\be\label{ORshell}
    \left.
    \mathcal O_R
    \right|_{\mathbf C=0}
    =
    \frac12R_{ab}R^{ab}
    -
    \frac14R^2\,.
\ee
We now compare the characteristic evolution with the curvature
evolution generated by the Ricci flow
\be
    \dot g_{ab}=-2R_{ab}\,.
\ee
The standard Ricci-flow identities are
\be
    \dot R_{ab}
    =
    \Delta R_{ab}
    +
    2R_{acbd}R^{cd}
    -
    2R_{ac}R^c{}_b
\ee
and
\be
    \dot R
    =
    \Delta R
    +
    2R_{ab}R^{ab}\,.
\ee
Since $\dot g^{ab}=2R^{ab}$, raising one index also contributes to
the evolution. Passing to the mixed Ricci tensor then removes the
quadratic Ricci term:
\be\label{Ricmix}
    \dot R^a{}_b
    =
    \Delta R^a{}_b
    +
    2R^a{}_{cbd}R^{cd}\,.
\ee
Differentiating
\be
    T^a{}_b
    =
    -R^a{}_b+\frac12R\,\delta^a_b
\ee
along the Ricci flow gives
\bea
    \dot T^a{}_b
    ={}&
    -\Delta R^a{}_b
    -
    2R^a{}_{cbd}R^{cd}
    \nonumber\\
    &+
    \frac12
    \left(
        \Delta R
        +
        2R_{cd}R^{cd}
    \right)\delta^a_b\,.
\label{Tgeom}
\eea
On the other hand, the characteristic equation gives
\be
    \dot{\mathbf T}
    =
    R\,\mathbf T
    +
    \mathcal O_R\,\mathbf 1
\ee
on the Einstein constraint surface. Using
Eq.~\eqref{ORshell}, one obtains
\be
    \dot T^a{}_b
    =
    -RR^a{}_b
    +
    \left(
        \frac12R_{cd}R^{cd}
        +
        \frac14R^2
    \right)\delta^a_b\,.
\ee
Equating the two expressions yields
\bea
    \mathfrak K^a{}_b
    \equiv{}&
    \Delta R^a{}_b
    +
    2R^a{}_{cbd}R^{cd}
    -
    RR^a{}_b
    \nonumber\\
    &-
    \left(
        \frac12\Delta R
        +
        \frac12R_{cd}R^{cd}
        -
        \frac14R^2
    \right)\delta^a_b
    =
    0\,,
\eea
which is Eq.~\eqref{firstobs} of the main text. Its trace is
\be
    \mathfrak K^a{}_a
    =
    -\frac12
    \left[
        (d-2)\Delta R
        +
        (d-4)
        \left(
            R_{ab}R^{ab}
            -
            \frac12R^2
        \right)
    \right].
\ee
Thus the tensorial condition reproduces the scalar obstruction
\eqref{scalarobs}; in four dimensions,
\be
    \Delta R=0\,.
\ee
\subsection{Higher-order compatibility}
\label{highercompat}
Let
\be
    \mathbf C
    =
    \mathbf G+\mathbf T
\ee
denote the Einstein constraint, and let $\mathcal V$ be the combined
characteristic vector field acting on $(g,\mathbf T)$. Exact
compatibility requires the surface $\mathbf C=0$ to be invariant
under $\mathcal V$.

For an analytic characteristic trajectory starting from an
Einstein--matter configuration, the first compatibility condition
at the initial point is
\be
    \left.
    \mathcal L_{\mathcal V}\mathbf C
    \right|_{\tau=0}
    =0\,,
    \qquad
    \left.\mathbf C\right|_{\tau=0}=0\,,
\ee
which is equivalent to
\be
    \left.\mathbf K\right|_{\tau=0}=0\,.
\ee
Successive conditions are obtained by repeated differentiation,
\be\label{apphier}
    \left.
    \mathcal L_{\mathcal V}^{\,n}\mathbf C
    \right|_{\tau=0}
    =0\,,
    \qquad n=1,2,\ldots\,.
\ee
The hierarchy has a simple interpretation: the first condition
requires tangency of the characteristic vector field to the Einstein
constraint surface at the initial configuration, while the higher
conditions prevent the characteristic trajectory from departing from
the constraint surface at successive orders in $\tau$. Thus
\be
    \boldsymbol{\mathfrak K}=0
\ee
is necessary but not sufficient in general. 
The trajectory-independent counterpart of this hierarchy is the
intrinsic invariance condition
\be
    \left.
    \mathcal L_{\mathcal V}\mathbf C
    \right|_{\mathbf C=0}
    =0
\ee
at every point of the constraint surface. This condition implies the
hierarchy above for every Einstein--matter initial configuration.
More generally, a closure
relation of the form
\be
    \mathcal L_{\mathcal V}\mathbf C
    =
    \mathcal A[\mathbf C]
\ee
for a local linear operator $\mathcal A$ makes the constraint surface
invariant and automatically implies the entire hierarchy
\eqref{apphier}.
\section{Explicit obstruction tests}
\label{apptests}
We give here the explicit calculations for the two examples used in
Section~\ref{obssec}.

\subsection{Reissner--Nordstr\"om}
\label{appRN}

Consider
\be
    \mathrm d s^2
    =
    -f(r)\,\mathrm d t^2
    +
    \frac{\mathrm d r^2}{f(r)}
    +
    r^2\mathrm d\Omega_2^2\,,
\ee
with
\be
    f(r)
    =
    1-\frac{2M}{r}
    +\frac{Q^2}{r^2}\,.
\ee
The mixed Ricci tensor is
\be
    R^a{}_b
    =
    \frac{Q^2}{r^4}
    \operatorname{diag}(-1,-1,1,1)\,,
\ee
so that
\be
    R=0\,,
    \qquad
    R_{ab}R^{ab}
    =
    \frac{4Q^4}{r^8}\,.
\ee
Since $R=0$, the four-dimensional scalar obstruction
$\Delta R=0$ is therefore satisfied identically.

For $R=0$, the full first obstruction reduces to
\be
    \mathfrak K^a{}_b
    =
    \Delta R^a{}_b
    +
    2R^a{}_{cbd}R^{cd}
    -
    \frac12R_{cd}R^{cd}\delta^a_b\,.
\ee
A direct evaluation gives
\be\label{RNcalc}
    \mathfrak K^a{}_b
    =
    \frac{4Q^2 f(r)}{r^6}
    \operatorname{diag}(-3,-1,2,2)
    \,.
\ee
The result is traceless, as required by $R=0$, but does not vanish
identically for $Q\neq0$. Hence the Reissner--Nordstr\"om geometry
fails the first tensorial compatibility condition despite satisfying
the scalar one.
\subsection{A cosmological higher-order obstruction}
\label{appFLRW}
Consider the spatially flat four-dimensional FLRW metric
\be
    \mathrm d s^2
    =
    -\mathrm d t^2
    +
    a(t)^2\,\mathrm d\mathbf x^2
\ee
with
\be\label{FLRWapp}
    a(t)=a_0t^{1/2}\,.
\ee
For this geometry
\be
    R=0\,,
\ee
and
\be
    R^a{}_b
    =
    \frac{1}{4t^2}
    \operatorname{diag}(-3,1,1,1)\,.
\ee
Direct substitution into the first obstruction gives
\be
    \mathfrak K^a{}_b=0
    \,.
\ee
Thus the characteristic vector field is tangent to the Einstein constraint surface at the initial configuration. To test the next order, let
\be
    \mathbf T(\tau)
    =
    \operatorname{diag}
    \left(
        u(\tau,t),
        v(\tau,t),
        v(\tau,t),
        v(\tau,t)
    \right).
\ee
Initially,
\be
    u_0=\frac{3}{4t^2}\,,
    \qquad
    v_0=-\frac{1}{4t^2}\,.
\ee
Substituting this eigenvalue structure into the characteristic
equations \eqref{Rchars} and expanding about $\tau=0$ gives
\bea
    u(\tau,t)
    &=
    \frac{3}{4t^2}
    +
    \frac{3\tau}{8t^4}
    +
    \frac{9\tau^2}{16t^6}
    +
    O(\tau^3)\,,
\\
    v(\tau,t)
    &=
    -\frac{1}{4t^2}
    +
    \frac{3\tau}{8t^4}
    -
    \frac{3\tau^2}{16t^6}
    +
    O(\tau^3)\,.
\eea
The metric characteristic preserves homogeneity and isotropy but
generates a nontrivial lapse,
\be
    \mathrm d s^2(\tau)
    =
    -N(t,\tau)^2\,\mathrm d t^2
    +
    A(t,\tau)^2\,\mathrm d\mathbf x^2\,,
\ee
with
\be
    N^2
    =
    1
    +
    \frac{3\tau}{2t^2}
    +
    \frac{3\tau^2}{4t^4}
    +
    O(\tau^3)
\ee
and
\be
    A^2
    =
    a_0^2t
    \left[
        1
        -
        \frac{\tau}{2t^2}
        -
        \frac{\tau^2}{4t^4}
        +
        O(\tau^3)
    \right].
\ee
Computing the Ricci tensor of the characteristic metric gives
\bea
    R^t{}_t[g(\tau)]
    ={}&
    -\frac{3}{4t^2}
    +
    \frac{3\tau}{8t^4}
    -
    \frac{57\tau^2}{8t^6}
    +
    O(\tau^3)\,,
\\
    R^i{}_j[g(\tau)]
    ={}&
    \left[
        \frac{1}{4t^2}
        +
        \frac{3\tau}{8t^4}
        -
        \frac{9\tau^2}{8t^6}
    \right]\delta^i_j
    +
    O(\tau^3)\,.
\eea
By contrast, the Einstein equations applied to the characteristic
stress tensor would require
\bea
    \left.R^t{}_t\right|_{\rm E}
    ={}&
    -\frac{3}{4t^2}
    +
    \frac{3\tau}{8t^4}
    -
    \frac{9\tau^2}{16t^6}
    +
    O(\tau^3)\,,
\\
    \left.R^i{}_j\right|_{\rm E}
    ={}&
    \left[
        \frac{1}{4t^2}
        +
        \frac{3\tau}{8t^4}
        +
        \frac{3\tau^2}{16t^6}
    \right]\delta^i_j
    +
    O(\tau^3)\,.
\eea
The two evolutions therefore agree through first order but differ at
second order:
\be
    R^t{}_t[g(\tau)]
    -
    \left.R^t{}_t\right|_{\rm E}
    =
    -\frac{105}{16}
    \frac{\tau^2}{t^6}
    +
    O(\tau^3)\,,
\ee
and
\be
    R^i{}_j[g(\tau)]
    -
    \left.R^i{}_j\right|_{\rm E}
    =
    -\frac{21}{16}
    \frac{\tau^2}{t^6}\,
    \delta^i_j
    +
    O(\tau^3)\,.
\ee
Hence
\be
    \mathbf C_0=0\,,
    \qquad
    \mathbf C_1=0\,,
    \qquad
    \mathbf C_2\neq0
    \,.
\ee
This provides an explicit example in which the complete first
obstruction vanishes, while the characteristic trajectory leaves the
Einstein constraint surface at second order in the deformation
parameter.

\bibliographystyle{JHEP}

\bibliography{masterP1}
\end{document}